\documentstyle[11pt,newpasp,twoside,epsf]{article}
\def\edcomment#1{\iffalse\marginpar{\raggedright\sl#1\/}\else\relax\fi}
\reversemarginpar

\begin{document}
\title{Viewing the Evolution of Massive Star Formation through FIR/Sub-mm/mm Eyes}
 \author{Lihong Yao}
\affil{Department of Astronomy \& Astrophysics, University of Toronto,
Toronto, ON M5S 3H8, Canada}

\author{E. R. Seaquist}
\affil{Department of Astronomy \& Astrophysics, University of Toronto,
Toronto, ON M5S 3H8, Canada}

\begin{abstract}
In this paper, we present an overview of our method of constructing a family of 
models for the far-infrared, sub-millimeter, and millimeter (FIR/sub-mm/mm) line 
emission of molecular and atomic gas surrounding massive star formation in starburst 
galaxies. We show the results of a case study, an expanding supershell centered 
around a massive star cluster with a particular set of input parameters and 
its application to nearby starburst galaxy M 82. This set of models can 
be used not only to interpret the observations of FIR/sub-mm/mm line emission
from molecular and atomic gas, but also to investigate the physical 
environment and the initial cloud conditions in massive star forming 
regions as well as the ages of the starbursts through simulations for a 
wide range of input parameters. Finally, we discuss limitations of our models,
and outline future work.
\end{abstract}

\section{Overview} 

A starburst phenomenon occurs when the star formation rate (SFR) is so rapid that 
it cannot be sustained for the lifetime of the galaxy. Giant molecular clouds (GMCs), 
especially the dense cores, are the places of active, massive star formation in 
galaxies. Starburst galaxies have impressive reservoirs of molecular gas at their 
centers to fuel the massive star formation. The effect of a newly born 
super star cluster (SSC) inside a GMC is to produce a hot bubble and a thin dense 
shell of neutral gas and dust swept up by the H II expansion, strong stellar winds, 
and repeated supernova (SN) explosions. Lying at the inner edges of the shells are 
the photodissociation regions (PDRs), the origin of much of the FIR/sub-mm/mm 
radiation from the interstellar medium (ISM) in starburst galaxies (see Figure 1). 
The bursts of massive star formation have a profound impact on the structure and 
evolution of their host galaxies by injecting large amounts of energy and mass 
into the ISM.

In the past, considerable efforts have been made on the observations of the 
FIR/sub-mm/mm line emission of the neutral gas in massive star forming regions 
of nearby starburst galaxies. Several models have been developed to interpret the
observed data (e.g. Mao et al. 2000; Wild et al. 1992). They have shown that the 
physical conditions (i.e. gas temperature, density, and FUV flux) are enhanced 
in starburst regions. The main drawback is that the extragalactic sources are 
far away, and hence higher resolution and sensitivity are required to map 
individual starburst regions in these galaxies. Theoretical starburst models 
can however synthesize the observations at any resolution and provide predictions 
and constraints for the expected gas behavior and conditions within star forming 
regions of galaxies. 

A successful model for the FIR/sub-mm/mm line emission of a starburst galaxy
must bring together in one package a wide range of physics including 
the dynamics of the bubble/shell structure around a young star cluster, 
the evolution of the stellar population, the fully time-dependent PDR chemistry, 
and finally, the method for solving the non-LTE line radiative transfer 
in molecular and atomic gas in star-forming regions. Few models, if any, 
have all these physical elements included at the same time. We introduce a new 
family of models that predict the properties of FIR/sub-mm/mm line emission 
from the evolving regions of massive star formation in a starburst galaxy, 
with particular emphasis on the neutral ISM (Yao et al., in preparation). The 
models follow the evolution of an ensemble of optically thick GMCs centrally 
illuminated by young evolving super star clusters (SSCs). A time-dependent 
treatment of the PDR chemistry is taken into account in the line radiative 
transfer modeling (Yao et al. 2006, and references therein). 

In this study, we will not address issues related to the triggering 
mechanisms of massive star formation, which are important in understanding how 
starburst galaxies form and evolve. However, by comparing our predicted 
line intensity ratios from ensembles of expanding shells with observational 
data of starburst galaxies, we hope to achieve a better understanding of 
(1) the current physical state of molecular and atomic gas surrounding 
massive star formation in starburst environments; (2) the initial conditions 
of the GMCs prior to star formation; (3) how the changes in star-formation related  
parameters (metallicity, radiation field strength, ambient pressure, and 
star-formation rate/efficiency) affect the properties and evolution of the 
neutral ISM in starburst galaxies; (4) how the CO-to-H$_2$ conversion factor $X$, 
a parameter that is used to determine the total molecular gas mass, changes as a 
starburst system evolves; and (5) the relationship between  a sequence of 
starbursts in a galaxy and the observed FIR/sub-mm/mm properties of 
the molecular and atomic gas to its age.

\section{Application to Nearby Starburst Galaxy M 82}

The starburst activity in M 82 (distance $\sim$ 3.25 Mpc) was likely triggered 
by tidal interaction with its companion M 81 beginning about 10$^8$ yr ago 
in the nucleus, and is currently propagating outward. The infrared luminosity 
of M 82 is about 4 $\times$ 10$^{10}$ L$_{\odot}$ arising mostly from the 
central 400 pc region, which has a stellar bar structure and currently has 
a high supernova rate of $\sim$0.05 - 0.1 yr$^{-1}$. The evolutionary scheme 
in M 82 remains under debate. The most commonly suggested ages of the starburst 
in the central region are 3 - 7 and 10 - 30 Myr (e.g. F\"{o}rster-Schreiber et al. 2003).

Using the expanding shell model described in the overview, our predictions of the
CO, HCN, and HCO$^+$ line intensity ratios agree with the molecular data for 
the central lobes (300 - 600 pc) for a shell with an age in the 3 - 7 Myr range 
(see Figure 1). This implies that the molecular torus is possibly a consequence 
of swept-up or compressed gas caused by massive star formation originating in the 
nucleus of M 82 such as those proposed by Carlstrom \& Kronberg (1991). However, 
it is important to realize the foregoing interpretation of the lobes as a ring 
or torus is not unique. A number of authors have argued that the molecular rings 
may be a product of Linblad resonance instabilities associated with the gravitational 
effects of the bar (e.g. Shen \& Lo 1996; Wills et al. 2000).   

\begin{figure}
\plottwo{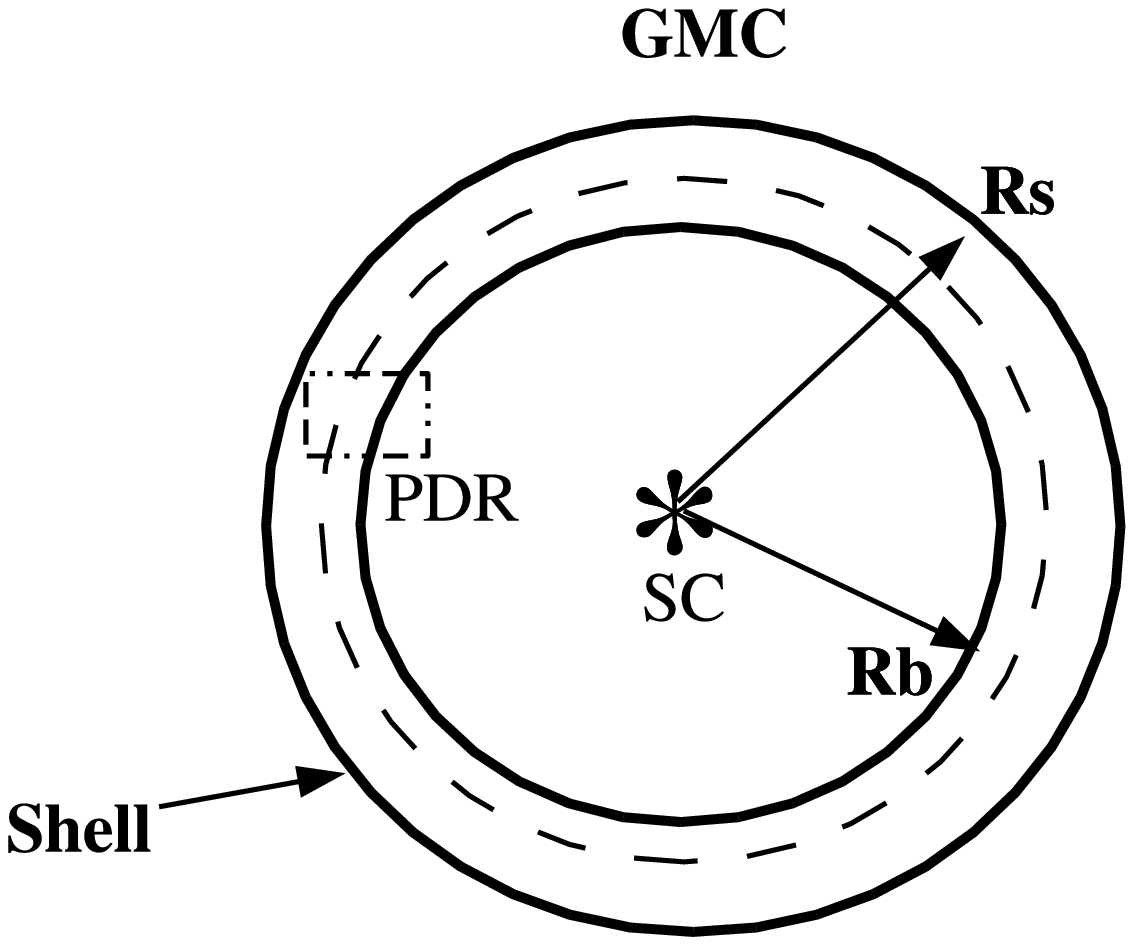}{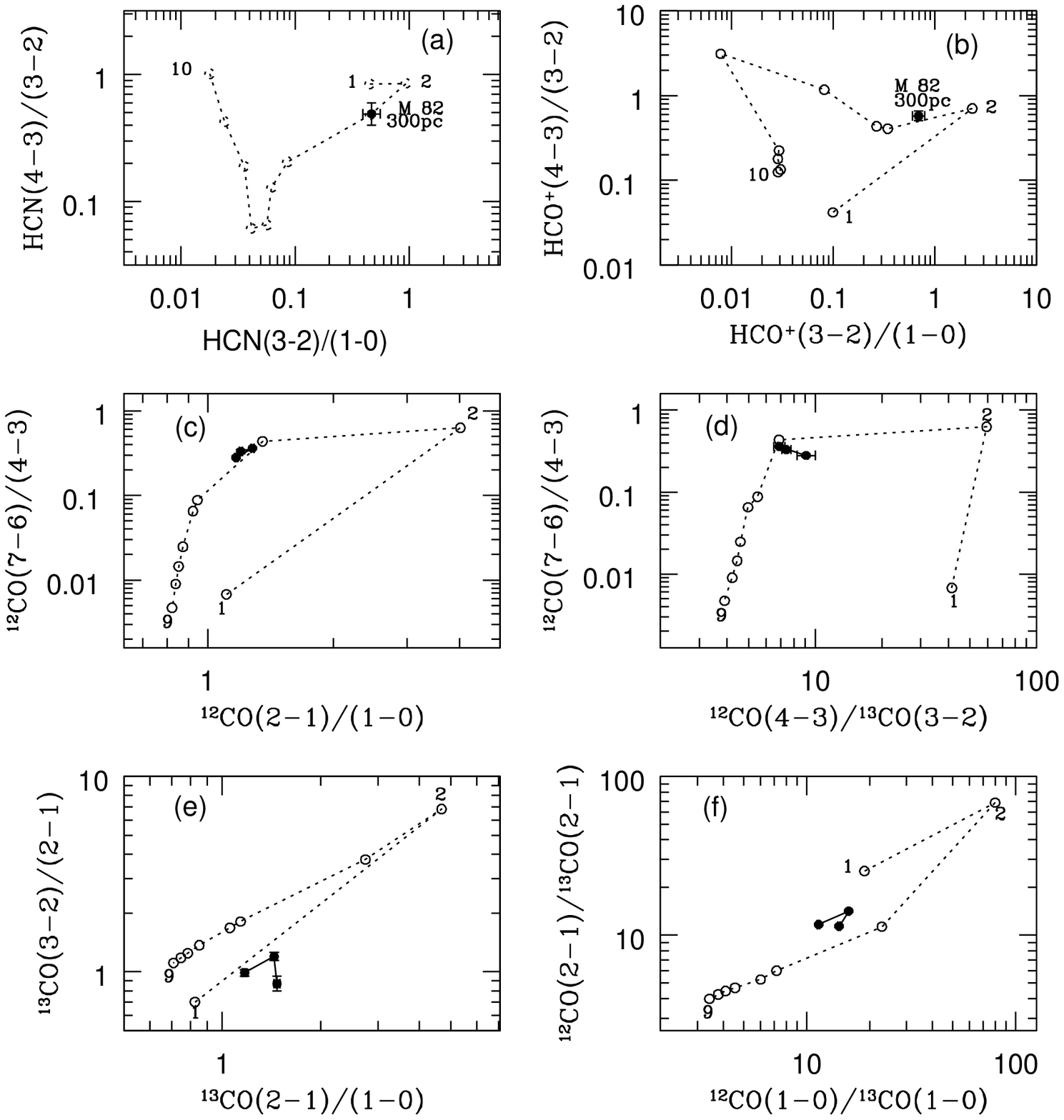}
\caption{Left: The schematic structure of an evolving GMC centrally illuminated by a 
compact young star cluster. Right: The ratio-ratio diagrams of the CO, HCN and HCO$^+$  
line intensities.}
\end{figure}

Our preliminary study also shows that the kinematic properties of the swept-up 
shell predicted by our model are in very good agreement with the measurements 
of the supershell centered around the bright SNR 41.9+58 in M 82 (e.g. Weiss et al. 
1999). This implies that the expanding supershell is created by strong 
stellar winds and supernova explosions from a young massive star cluster 
($\sim$ 2.5 $\times$ 10$^6$M$_{\odot}$) located at its center (see Table 1).

\begin{table}
\caption{Characteristics of The Expanding Supershell in M 82. \label{tbl-1}}
\begin{tabular}{lll} \hline\hline
{\b Parameter} & {\b Observation} & {\b Model} \\
Radius (pc) & 65.0 & 65.0 \\
Age (Myr)   & 1.0  &  1.0 \\
Expansion velocity (km s$^{-1}$) & 45 & 45 \\ 
Total H$_2$ molecular gas mass ($\times$ 10$^6$ M$_{\odot}$) & 8.0 & 7.6 \\
Kinetic Energy ($\times$ 10$^{53}$ ergs) & 1.6 & 1.5 \\
Total stellar mass in the center cluster ($\times$ 10$^6$ M$_{\odot}$) & $\ldots$ & 2.5  \\
Total number of O stars ($\ge$ 40 M$_{\odot}$) & $\ldots$   & 1700 \\
Total Mechanical Energy ($\times$ 10$^{54}$ ergs) & $\ldots$ & 1.7  \\ \hline \hline           
\end{tabular}
\end{table}

We also modeled the $^{12}$CO line intensity ratios of high-$J$ transitions to 
the 1-0 transition, and the [C I], [C II] line intensity ratios for the expanding 
supershell centered around SNR 41.9+58 in M 82 (Figure 2). The predictions can be 
used for comparison with future observations, and also to constrain the physical 
conditions of the gas in the shell. The model line ratios which are greater 
than unity at $t$ $<$ 8 Myr imply that CO is optically thin in the expanding
supershell. Therefore, it is better to look at the high CO transitions ($J$ $>$ 3) 
in this supershell. Observational data for this region are not available in multiple 
lines because of the high resolution needed. However, Seaquist et al. (2006) 
shows some evidence for enhancement in $r_{61}$ in the supershell at 7$\arcsec$ 
resolution. The ratio is not as high as the model prediction. However, the data 
do not completely isolate the supershell emission.

\section{Conclusion and Future Work}

Our models yield FIR/sub-mm/mm molecular and atomic line intensities and radii 
for shells based on initial cloud conditions appropriate for those in M 82. The 
models suggest that the neutral ISM of the central star-forming region is a product 
of fragments of the evolving shells. The good agreement between our model and 
observed molecular line intensities as well as the observed sizes and motions of 
some shell structures indicates that the set of models we have developed can be 
used to interpret star forming activity in other starburst systems too.

The limitations of our models are that we do not include the effects of magnetic 
fields, the interactions between shells or between shell and cloud, the effects of 
clumpy clouds and the non-uniform ambient ISM, and the effects of non-spherical 
cloud geometries and cloud spatial distributions.

A number of issues arise from our preliminary study of the expanding supershell 
in M 82. It is not clear whether our results support a physical association between 
the supershell and the bright SNR 41.9+58 near its center. If the SNR were within 
or near the SSC, there is an issue whether there is sufficient gas remaining 
to form an SNR after the action of the previous winds and SNe. 
(2). The SSC responsible for the formation of the supershell might also have 
provided the stellar mass for the several hundred solar mass black hole detected 
by Chandra X-ray observations near its center. Theories for the formation of this 
black hole include the collapse of a ``hyperstar'' formed by the coalescence of 
many normal stars, or the direct merger of stellar mass black holes. 
The SSC is adequately endowed with sufficient mass since there would have 
been 1,700 O stars, each with mass greater than or equal to about 40 M$_{\sun}$.
(3). The picture we developed for the M 82 molecular ring is similar to that of 
Carlstrom \& Kronberg (1991), in which the primary ring or torus surrounding 
the nucleus is the product of star formation in the nuclear region. Though such 
a model yields the observed line ratios for molecular transitions, the atomic line 
ratios calculated by our model do not fit the observed data as well, and suggest 
a much older shell. This is possibly because the atomic line data 
have such low resolution that they correspond to a much larger region ($>$ 1 kpc). 
A variety of modeling parameters and comparisons with high resolution observational 
data need to be considered to yield a more accurate picture and more precise ages 
of the starbursts in M 82.

\acknowledgements 
This research was supported by a research grant from the Natural Sciences and 
Engineering Research Council of Canada (to E. R. S.), and additional fellowship 
support from the Department of Astronomy and Astrophysics at the University of 
Toronto, including a Reinhardt Travel Fellowship.  I thank Peter van Hoof for 
comments.

\begin{figure}
\vspace{-1.2in}
\plotone{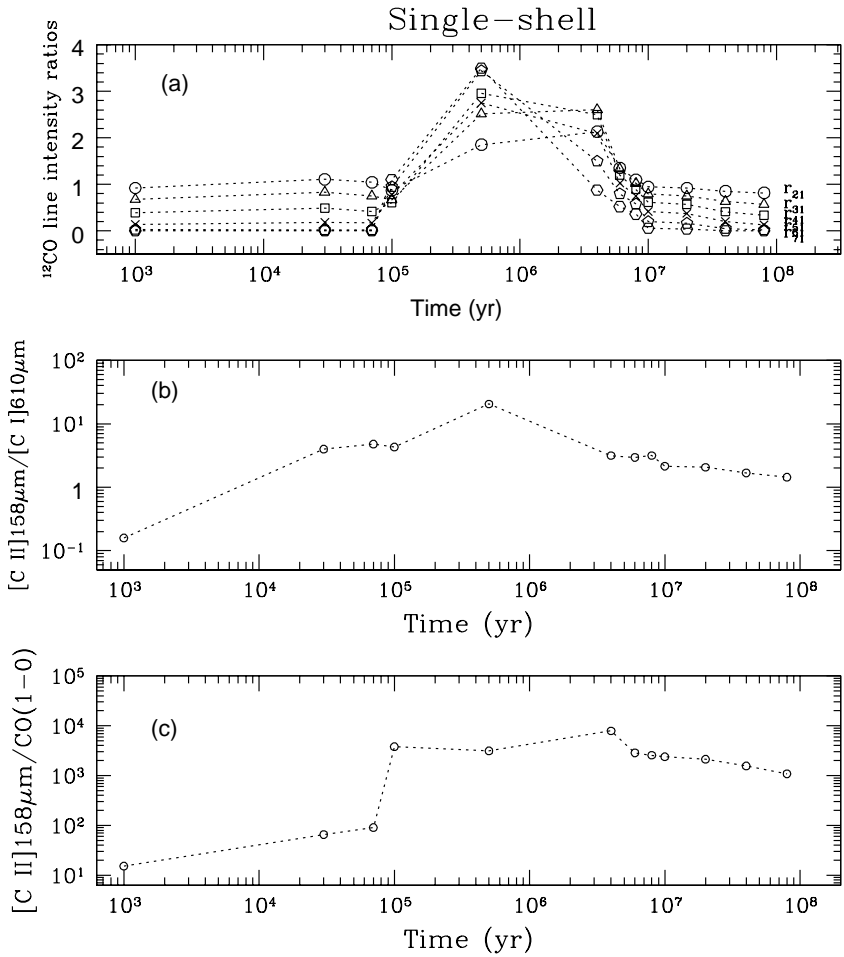}
\caption{Theoretical modeling of $^{12}$CO line intensity ratios of high-$J$ transitions to 1-0 transition, and [C I], [C II] line intensity ratios for the expanding supershell in M 82.}
\end{figure}

\end{document}